%% file: Qcolour_amp.tex
\documentclass[11pt,epsf]{article}
 \usepackage{amsmath}
 \usepackage{graphicx}
 \usepackage[merge,numbers,compress]{natbib}
 \usepackage[T1]{fontenc}
 \usepackage{booktabs}
 \usepackage{xcolor} 
 \usepackage{xspace}
 \usepackage{dcolumn}
 \usepackage{placeins}
 \usepackage{amssymb}
 \usepackage[colorlinks=true,citecolor=blue!50!black,linkcolor=black]{hyperref}
 \usepackage[font=small,labelfont=bf]{caption}
 \usepackage
 [subrefformat=parens,position=top,skip=-15pt,margin=15pt,justification=justified,singlelinecheck=false]
 {subcaption}
 \usepackage{array,multirow}
 \usepackage{braket}
 \usepackage{makecell}
 \usepackage{dsfont}
 \usepackage{caption}
 \usepackage{subcaption}
 \usepackage{orcidlink}
 \usepackage{soul}

\newcommand{\id}{\mathds{1}}

\newcommand{\sutwo}{\mathfrak{su}(2)}
\newcommand{\suNc}{\mathfrak{su}(N_c)}
\newcommand{\Qsutwo}{Q_{\mathfrak{su}(2)}}

\input{macros}

\begin{document}

\title{\hfill ~\\[-30mm]
\phantom{h} \hfill\mbox{\small FR-PHENO-2025-007}
\\
\phantom{h} \hfill\mbox{\small IPPP/25/34}
\\[0.5cm]
\vspace{13mm}   \textbf{Quantum simulation of scattering amplitudes and interferences in perturbative QCD}}

\date{}
\author{
Herschel A.\ Chawdhry$^{1\,}$\footnote{E-mail:  \texttt{hchawdhry@fsu.edu}} \orcidlink{0000-0002-2901-3353},
Mathieu Pellen$^{2\,}$\footnote{E-mail:  \texttt{mathieu.pellen@physik.uni-freiburg.de}} \orcidlink{0000-0001-5324-2765}, and
Simon Williams$^{3\,}$\footnote{E-mail:  \texttt{simon.j.williams@durham.ac.uk}} \orcidlink{0000-0001-8540-0780}
\\[9mm]
{\small\it $^1$ Department of Physics, Florida State University,} \\ 
{\small\it 77 Chieftan Way, Tallahassee FL 32306, USA} \\[3mm] %
{\small\it $^2$ Albert-Ludwigs-Universit\"at Freiburg, Physikalisches Institut,} \\ %
{\small\it Hermann-Herder-Stra\ss e 3, D-79104 Freiburg, Germany}\\[3mm]
{\small\it $^3$ Institute for Particle Physics Phenomenology,} \\ %
{\small\it Durham University, Durham DH1 3LE, UK}\\[3mm]
        }
\maketitle

\begin{abstract}
\noindent
A flagship application of quantum computers is the simulation of other quantum systems, including quantum field theories.
In this article, we show how quantum computers can be employed to naturally calculate Feynman diagrams and their interferences
in Quantum Chromodynamics (QCD).
We simulate the colour parts of the interactions directly on the quantum computer, while the kinematic parts are for now pre-computed classically.
For processes where some of the external particles are identical, we find the first hints of a potential quantum advantage.
We validate our techniques using simulated quantum computers.
Furthermore, for toy examples we also demonstrate our algorithms on a 56-qubit trapped-ion quantum computer.
The work constitutes a further key step towards a full quantum simulation of generic perturbative QCD processes.
\end{abstract}
\thispagestyle{empty}
\vfill

\newpage

\tableofcontents

\section{Introduction}
Scattering amplitudes are of core importance in high-energy physics, due to their central role in predicting the scattering, production, and decay of fundamental particles.
These amplitudes are defined in the framework of Quantum Field Theory (QFT), and their precise calculation is vital for relating our fundamental models of Nature to the experimental measurements performed at particle colliders such as the Large Hadron Collider (LHC).
This enables stringent validations of the Standard Model of particle physics, as well as increasing our ability to detect potential new particles or phenomena beyond the Standard Model.
Since the LHC collides protons, composed of quarks and gluons, one relies in large part on calculations in Quantum Chromodynamics (QCD), the QFT that describes the strong interactions of quarks and gluons.

The high-energy regime of QCD is of special interest, because it is extensively probed by LHC events and is therefore of most direct relevance for studying the properties of Nature at the shortest distance scales presently accessible to experiments.
For these energies, far above the fundamental QCD scale $\Lambda_{\textrm{QCD}} \sim 200 \textrm{MeV}$, a direct lattice-based simulation of QCD is likely to remain well beyond reach for the forseeable future due to the large number of lattice sites that it would require.
At the same time, at these energies the QCD coupling constant $\alphas$ becomes small, allowing scattering amplitudes and their associated physical observables to be calculated as perturbative series in $\alphas$.
Perturbative QCD calculations thus play a central role in predicting the outcomes of high-energy collisions~\cite{Huss:2025nlt}.

Perturbative QCD calculations are computationally challenging, which limits our ability to predict the results of experiments, infer the properties of known fundamental particles, or detect the existence of new ones.
New computational techniques are thus constantly being sought.
In view of recent advances in quantum computing hardware technology~\cite{PhysRevX.13.041052,Kim:2023bwr,GoogleQuantumAIandCollaborators:2024efv,Grinkemeyer:2024ofj,AghaeeRad:2025lqu,larsen2025integrated}, it is natural to consider whether quantum computers could be beneficial for calculating perturbative QCD scattering amplitudes.
The investigation of this question is the purpose of this article, and indeed we find the first hints of a potential quantum advantage, \textit{i.e.} indications that quantum computers could outperform classical computers in this domain.

Perturbative QCD observables, such as scattering cross-sections and particle decay rates, are calculated from the modulus-square $|\mathcal{A}|^2$ of the amplitude $\mathcal{A}$, which can in turn be written as a sum over Feynman diagrams:
\begin{equation}\label{eq:ampl_sum_diagrams}
\mathcal{A} = \sum_n C_n K_n.
\end{equation}
Here for later convenience, each Feynman diagram $F_n$ has been decomposed into a product of a colour factor $C_n$ and a kinematic factor $K_n$.\footnote{Quartic gluon interactions cannot be factorised in this way, but their contributions can be expanded as a sum of terms, each factorised into a product of a colour factor and a kinematic factor.}
Since quantum computers operate by manipulating superpositions of states, they are natural candidates for simulating perturbative QCD processes.
Indeed, when performing a measurement of the quantum computer, the probability of a given outcome is determined by the modulus-square of the relevant component of the wavefunction of the quantum computer.
These commonalities will be exploited in several ways in this article: first to directly simulate the colour parts $C_n$ of individual Feynman diagrams, and later to obtain the quantum-coherent sums and interferences of multiple Feynman diagrams.

The possibility of using quantum computers to simulate quantum systems has previously been studied in various contexts, with examples ranging from chemistry~\cite{ChemistryRev,RevModPhys92015003} and condensed matter~\cite{RevModPhys.86.153,qute.201900052,Zhou:2021kdl,Mallick:2024slg} to non-perturbative QFT simulations~\cite{Jordan:2011ci,Jordan:2012xnu,Martinez:2016yna,Klco:2018zqz,Banuls:2019bmf,Davoudi:2020yln,Klco:2021lap,Carena:2022kpg,Rigobello:2023ype,Zemlevskiy:2024vxt,Anderson:2024kfj,Cochran:2024rwe,Gonzalez-Cuadra:2024xul}.
However, there is a significant difference between those applications and ours: most such proposals seek to use a known Hamiltonian to perform the unitary evolution of a quantum system, usually by means of Trotterisation~\cite{trotter1959product,suzuki1976generalized}, whereas in perturbative QFT calculations one seeks to calculate the (Hermitian, but not unitary) transition matrix between specified external states.

The first step towards a quantum simulation of generic perturbative QCD processes was taken in Ref.~\cite{Chawdhry:2023jks}, which introduced two fundamental quantum circuits, denoted $Q$ and $G$, to simulate the colour parts of the quark-gluon and triple-gluon interactions, respectively.
In this way, the circuits provide building blocks for the colour factors $C_n$ appearing in Eq.~\eqref{eq:ampl_sum_diagrams}.
In the present article we start by employing $Q$ and $G$ gates to simulate the colour parts of individual Feynman diagrams (Sec.~\ref{sec:ampl_level_tracing}), in a refinement of the colour-tracing algorithm of Ref.~\cite{Chawdhry:2023jks}.
Next, in Sec.~\ref{sec:interference} we present techniques to simulate multiple Feynman diagrams on a quantum computer and obtain their interference via the measurement process.
We then in Sec.~\ref{sec:permutation} consider the special case of Feynman diagrams related to one another by the interchange of identical external particles, and show efficient ways to obtain the sums and interferences of these diagrams.
To validate these new algorithms, we implement them on idealised noiseless quantum emulators as well as on the physical 56-qubit trapped-ion quantum computer H2-1 made by {\sc Quantinuum}.
The results from the noiseless emulators are presented alongside the relevant methods in Secs.~\ref{sec:interference} and~\ref{sec:permutation}, while for the demonstrations on the physical device we simplify to a 2-colour toy model of QCD and so we present the latter results separately in Sec.~\ref{sec:physical_devices}.

Finally, we highlight that our work fits into a broader context of proposed quantum-computing applications to high-energy physics which have emerged in the last few years~\cite{Bauer:2022hpo,Gray:2022fou,Delgado:2022tpc,Brown:2023llg,DiMeglio:2023nsa}, ranging from quantum integration~\cite{Agliardi:2022ghn,deLejarza:2023qxk,Cruz-Martinez:2023vgs,deLejarza:2024scm,Williams:2025hza} and quantum Monte Carlo~\cite{Bravo-Prieto:2021ehz,Robbiati:2023mbk,Lee:2025lzz}, to parton showers including interference effects~\cite{Bauer:2019qxa,Bepari:2020xqi,Bepari:2021kwv,Deliyannis:2022uyh,Gustafson:2022dsq,Bauer:2023ujy}, and also parton-distribution functions~\cite{Perez-Salinas:2020nem,Li:2021kcs} and loop-integrals~\cite{Ramirez-Uribe:2021ubp,Clemente:2022nll,deLejarza:2024pgk,Ramirez-Uribe:2024wua}.
The present article thus continues the growing effort to harness the potential of quantum computing in addressing complex problems in high-energy physics, aiming to uncover new applications and computational strategies within this domain.

\section{Amplitude-level colour simulation}\label{sec:ampl_level_tracing}
In general, the amplitude $\mathcal{A}$ and colour factors $C_n$ in Eq.~\eqref{eq:ampl_sum_diagrams} will carry free colour indices representing the colours of external particles, and so we can explicitly write
\begin{equation}\label{eq:ampl_sum_diagrams_explicit_indices}
\mathcal{A}^{a_1 a_2 \ldots}_{i_1 i_2 \ldots , j_1 j_2 \ldots} = \sum_n C^{a_1 a_2 \ldots}_{n, i_1 i_2 \ldots , j_1 j_2 \ldots} K_n.
\end{equation}
Here for each external gluon in the process there is an index $a_m$ in the adjoint representation of the QCD gauge group algebra $\suNc$, where $N_c=3$ is the number of quark colours, and for each external quark or external antiquark there is a fundamental-representation index $i_m$ or antifundamental-representation index $j_m$ as appropriate.
Some of the colour factors $C_n$ may have additional colour indices corresponding to internal particles, but such indices are contracted and so are not displayed in Eq.~\eqref{eq:ampl_sum_diagrams_explicit_indices}.
Our aim in this section is to encode the colour factor $C^{a_1 a_2 \ldots}_{n, i_1 i_2 \ldots , j_1 j_2 \ldots}$ of a single diagram $F_n$ into the wavefunction of the quantum computer.
Such a representation will allow us to directly obtain the colour-summed squared colour factor $C^2_n$ for our chosen diagram by performing a measurement on the quantum computer, but it will also lay the groundwork for obtaining interferences of multiple diagrams in the later sections.

We begin by summarising the behaviour of the $Q$ and $G$ gates from Ref.~\cite{Chawdhry:2023jks}, and then present a refined algorithm to combine these gates to produce the colour factor $C_n$ of any chosen Feynman diagram.
We represent the colour state of a particle by using a quantum computer register comprising several qubits.
For example, the $N_c^2-1=8$ colour basis states of a gluon can be represented by the $2^3=8$ basis states of a 3-qubit register.
The $Q$ and $G$ gates act upon these registers to simulate, respectively, the fundamental-representation generators $T^a_{ij}$ and the structure functions $f^{abc}$ of $\suNc$ in the following way.%
\footnote{As is common practice, the normalisation of the fundamental-representation generators in this article is chosen such that $T_F=\frac{1}{2}$ in the identity $\textrm{Tr}(T^a T^b)=T_F\delta^{ab}$.}
Given a register $q$ representing a quark in colour basis state $\ket{j}$, with $j \in \{1, \ldots, N_c\}$, and a register $g$ representing a gluon in colour basis state $\ket{a}$, with $a \in \{1, \ldots, N_c^2-1\}$, the $Q$ gate performs the transformation
\begin{equation}\label{eq:Q_gate_behaviour}
Q\ket{a}_g\ket{j}_q\ket{\Omega}_\mathcal{U} = \sum_{i=1}^{N_c} T^a_{ij} \ket{a}_g\ket{i}_q\ket{\Omega}_\mathcal{U} + \left(\textrm{terms orthogonal to } \ket{\Omega}_\mathcal{U} \right) .
\end{equation}
Here $\mathcal{U}$ is a small ancilla register whose inclusion allows for the simulation of non-unitary operators like $T^a_{ij}$ and $f^{abc}$, as further explained in Ref.~\cite{Chawdhry:2023jks}.
We use the notation $\ket{\Omega}_r$ to represent a reference state of register $r$ in which all qubits are in the $\ket{0}$ state.
Similarly to Eq.~\eqref{eq:Q_gate_behaviour}, the $G$ gate acts on three gluon registers as follows:
\begin{equation}\label{eq:G_gate_behaviour}
G \ket{a}_{g_1}\ket{b}_{g_2}\ket{c}_{g_3}\ket{\Omega}_{\mathcal{U}} = f^{abc} \ket{a}_{g_1}\ket{b}_{g_2}\ket{c}_{g_3}\ket{\Omega}_{\mathcal{U}} + \left(\textrm{terms orthogonal to } \ket{\Omega}_\mathcal{U} \right).
\end{equation}
Full details regarding the construction and implementation of the $Q$ and $G$ gates can be found in Ref.~\cite{Chawdhry:2023jks}.
We highlight that when combining multiple $Q$ and $G$ gates to simulate one or more Feynman diagrams, a single $\mathcal{U}$ register suffices, and the number of qubits it must contain is logarithmic in the number of vertices of any one Feynman diagram.

By applying a suitable sequence of $Q$ and $G$ gates, it is straight-forward to implement the colour part of a Feynman diagram as specified by the Feynman rules of QCD.
Indeed, in Ref.~\cite{Chawdhry:2023jks} we provided an explicit algorithm to do so and hence obtain the colour trace for a particular Feynman diagram.
However, that trace algorithm required all colour indices to be contracted, and therefore implicitly performed colour traces at the squared-amplitude level.
In the remainder of this section, we introduce an improved tracing algorithm that instead operates at the amplitude level.
The advantages of this approach are three-fold.
Firstly, it halves the circuit depth (\textit{i.e.} the number of sequential operations in the circuit), since an unsquared Feynman diagram has half as many factors of $T^a_{ij}$ and $f^{abc}$ as a squared diagram, and therefore requires half as many $Q$ and $G$ gates to simulate it.
Secondly, it reduces the number of shots (\textit{i.e.}\ repeated runs) required on the quantum device to attain a given level of precision.
Thirdly, and most importantly, it lays the groundwork for allowing multiple Feynman diagrams to be coherently summed at the amplitude level by the quantum computer, which we will exploit in Sections~\ref{sec:interference} and~\ref{sec:permutation} to calculate interferences of diagrams.

\begin{figure}[]
\center
        \begin{subfigure}{0.49\textwidth}
        \centering
                 \includegraphics[width=0.65\textwidth]{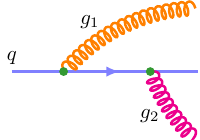}
                 \vspace{1.5cm}
                 \vspace{0.5cm}
                 \subcaption{\centering Example Feynman diagram}\label{fig:example_FeynmanDiagram}
        \end{subfigure}
\hfill
        \begin{subfigure}{0.49\textwidth}
        \includegraphics[width=0.9\textwidth]{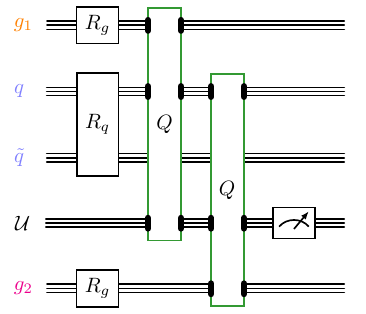}
\vspace{0.5cm}
                 \subcaption{\centering Corresponding quantum circuit}\label{fig:example_circuit}
        \end{subfigure}
        \vspace{0.5cm}
        \caption{\label{fig:example_FeynmanDiagramAndCircuit}%
Illustrative example Feynman diagram (left) and a quantum computing circuit to simulate it (right).
Qubits are grouped into registers, shown as triple horizontal lines.
Each gluon is represented by a register as indicated.
The quark line is represented by a pair of registers $q$ and $\tilde{q}$ to encode its fundamental and antifundamental colour index respectively.
There is also an ancilla register $\mathcal{U}$ for reasons explained in the main text.
Boxes represent operations acting on one or more registers.
Each $Q$ operation simulates a quark-gluon vertex by acting on a subset of registers identified by black dots.
The meter icon represents measurement.
}
\end{figure}

Before laying out our refined colour-tracing algorithm in full generality, we will start with an illustrative example.
We will consider the Feynman diagram in Fig.~\ref{fig:example_FeynmanDiagram}.
Its colour factor, according to the Feynman rules of QCD, is
\begin{equation}\label{eq:example_colour_factor}
\sum_l T^b_{il} T^a_{lj},
\end{equation}
where the adjoint indices $a$ and $b$ are associated, respectively, to the gluons $g_1$ and $g_2$, while the indices $j,l,i$ are associated, respectively, to the initial, intermediate, and final colour states of the quark.
This colour factor can be obtained by simulating the Feynman diagram using the quantum computer circuit in Fig.~\ref{fig:example_circuit}, whose construction and operation will now be explained step by step.
The circuit has 2 gluon registers, one for each gluon in Fig~\ref{fig:example_FeynmanDiagram}.
There is also a pair of quark registers $q$ and $\tilde{q}$, to represent the fundamental and antifundamental external index of the quark line, respectively.
In addition, there is a register $\mathcal{U}$ as discussed above.
Initially, each register $r$ is in the state $\ket{\Omega}_r$, and so the state of the quantum computer is
\begin{equation}
\ket{\Omega}_{g_1}\ket{\Omega}_{g_2}\ket{\Omega}_{q}\ket{\Omega}_{\tilde{q}}\ket{\Omega}_{\mathcal{U}}.
\end{equation}
We start by placing each gluon register $g$ into an equal superposition of colour states by applying a rotation operator $R_g$ designed such that
\begin{equation}\label{eq:R_g}
R_g \ket{\Omega}_g = \sum_{a=1}^8 \frac{1}{\sqrt{8}} \ket{a}_g.
\end{equation}
We also apply a rotation to the $q$ and $\tilde{q}$ registers using an operator $R_q$ designed such that
\begin{equation}\label{eq:R_q}
R_q \ket{\Omega}_q \ket{\Omega}_{\tilde{q}} = \sum_{j=1}^3 \frac{1}{\sqrt{3}} \ket{j}_q \ket{j}_{\tilde{q}}.
\end{equation}
Explicit constructions for $R_g$ and $R_q$ are given in Appendix A of Ref.~\cite{Chawdhry:2023jks}.
Thus, after applying the $R_g$ and $R_q$ gates appearing in Fig.~\ref{fig:example_circuit}, the state of the quantum computer is
\begin{equation}
\frac{1}{8\sqrt{3}} \sum_{
\substack{
a,b \in \{1, ..., 8\}\\
j \in \{1,2,3\}
}
} \ket{a}_{g_1} \ket{b}_{g_2} \ket{j}_{q} \ket{j}_{\tilde{q}} \ket{\Omega}_\mathcal{U}.
\end{equation}
We then apply the $Q$ gate in Eq.~\eqref{eq:Q_gate_behaviour} to registers $q$, $g_1$, and $\mathcal{U}$ to simulate the first gluon emission vertex in Fig.~\ref{fig:example_FeynmanDiagram}.
This places the quantum computer into a state
\begin{equation}
\frac{1}{8\sqrt{3}} \sum_{
\substack{
a,b \in \{1, ..., 8\}\\
j,l \in \{1,2,3\}
}
} T^a_{lj} \ket{a}_{g_1} \ket{b}_{g_2} \ket{l}_{q} \ket{j}_{\tilde{q}} \ket{\Omega}_\mathcal{U} + \left(\textrm{terms orthogonal to } \ket{\Omega}_\mathcal{U} \right) .
\end{equation}
Next, we apply another $Q$ gate, this time to registers $q$, $g_2$, and $\mathcal{U}$, to simulate the second gluon emission vertex in Fig.~\ref{fig:example_FeynmanDiagram}.
This places the quantum computer into a state
\begin{equation}\label{eq:example_final_state}
\frac{1}{8\sqrt{3}} \sum_{
\substack{
a,b \in \{1, ..., 8\}\\
i,j,l \in \{1,2,3\}
}
} T^b_{il} T^a_{lj} \ket{a}_{g_1} \ket{b}_{g_2} \ket{i}_{q} \ket{j}_{\tilde{q}} \ket{\Omega}_\mathcal{U} + \left(\textrm{terms orthogonal to } \ket{\Omega}_\mathcal{U} \right) .
\end{equation}
It can be seen that in this state, the coefficient of $\ket{\Omega}_\mathcal{U}$ neatly encodes the colour factor of the Feynman diagram.
We highlight that in this state, just as in Eq.~\eqref{eq:example_colour_factor}, the dummy index $l$ has been summed over.
The free indices $i,j,a,b$, on the other hand, remain encoded in the states of the registers $q$, $\tilde{q}$, $g_1$, $g_2$ respectively, which will turn out to be convenient in Secs.~\ref{sec:interference} and~\ref{sec:permutation} when we sum multiple Feynman diagrams and obtain their interferences.
It can also be seen that if the $\mathcal{U}$ register is measured, the probability of obtaining the reference state $\ket{\Omega}_\mathcal{U}$ is
\begin{equation}
P(\ket{\Omega}_\mathcal{U}) =
\frac{1}{192} \sum_{
\substack{
a,b \in \{1, ..., 8\}\\
i,j,l,m \in \{1,2,3\}
}
} T^a_{il} T^b_{lj} T^b_{jm} T^a_{mi},
\end{equation}
which is, up to a normalisation, the colour-averaged squared colour factor for the diagram in Fig.~\ref{fig:example_FeynmanDiagram}.

We can now generalise to a generic Feynman diagram $F_n$ containing $N_{g_I}$ internal gluons, $N_{g_E}$ external gluons, $N_{q_I}$ closed quark loops, and $N_{q_E}$ open quark lines.
For each gluon, we require a register containing 3 qubits, as explained above.
For each quark line, we require a pair of 2-qubit registers $q$ and $\tilde{q}$, which will encode respectively the fundamental and antifundamental index of the quark line.
We also require an ancilla register $\mathcal{U}$ comprising $\lceil \log_2 (N_V) \rceil$ qubits, where $N_V$ is the number of vertices in the Feynman diagram.
The colour part of $F_n$ can then be simulated using the following procedure:
\begin{enumerate}
 \item Initialise each register $r$ into the state $\ket{\Omega}_r \equiv \ket{0\dots 0}_r$.
 \item Apply $R_g$ to each gluon register.
 \item For each quark line, apply $R_q$ to the associated $q$ and $\tilde{q}$ registers.
 \item\label{step:applyQgates} For each quark-gluon interaction vertex, apply a $Q$ gate to the quark register $q$ and gluon register $g$ that correspond to the quark and gluon at that vertex.
 \item\label{step:applyGgates} For each triple-gluon interaction, apply a $G$ gate to the 3 corresponding gluon registers.
 \item\label{step:invRotateGluons} For each internal gluon, apply an $R_g^{-1}$ gate to the corresponding gluon register.
 \item\label{step:invRotateQuarks} For each internal quark line, apply an $R_q^{-1}$ gate to the corresponding pair of quark registers, $q$ and~$\tilde{q}$.
  \item\label{step:measurement} Measure the unitarisation register $\mathcal{U}$ and the registers associated with internal particles $I$, but not the registers associated with external particles $E$.
\end{enumerate}
This procedure is similar to that of Ref.~\cite{Chawdhry:2023jks}, but has three important refinements so that it operates at the amplitude level and can be used later for calculating interferences. Firstly, steps~\ref{step:applyQgates} and~\ref{step:applyGgates} are performed for a Feynman diagram instead of its square. Secondly, steps~\ref{step:invRotateGluons} and~\ref{step:invRotateQuarks} apply only to internal particles. Thirdly, the qubits associated with external particles are not measured.

Defining a shorthand $\ket{\Omega}_{r_1 + r_2 + \ldots}$ to indicate a state where registers $r_1, r_2, \ldots$ are each in a reference state $\ket{0\ldots 0}$, the final state $\ket{\Psi}$ of the quantum computer at the time of the measurement is, analogously to Eq.~\eqref{eq:example_final_state},
\begin{equation}\label{eq:amplitude_output_state}
\ket{\Psi} = \frac{1}{\mathcal{N}} \ket{\xi_n}_E \ket{\Omega}_{I + \mathcal{U}}
+ \left( \textrm{terms orthogonal to } \ket{\Omega}_{I + \mathcal{U}} \right),
\end{equation}
where $\ket{\xi_n}$ encodes the desired colour factor $C_n$ of the chosen Feynman diagram $F_n$ as:
\begin{equation}\label{eq:vector_xi}
\ket{\xi_n} = \sum_{
\substack{
a_1, a_2, \ldots \in \{1, \ldots, N_c^2-1\}\\
i_1, i_2, \ldots \in \{1, \ldots, N_c\}\\
j_1, j_2, \ldots \in \{1, \ldots, N_c\}
}
} C^{a_1 a_2 \ldots}_{n, i_1 i_2 \ldots, j_1 j_2 \ldots} \bigl(\ket{a_1}\ket{a_2}\ldots\bigr) \bigl(\ket{i_1}\ket{i_2}\ldots\bigr) \bigl(\ket{j_1}\ket{j_2}\ldots\bigr).
\end{equation}
Here, just as in Eq.~\eqref{eq:ampl_sum_diagrams_explicit_indices}, the indices $a_k, i_k, j_k$ are associated only with external particles.
In Eq.~\eqref{eq:amplitude_output_state}, we have introduced a normalisation factor
\begin{equation}\label{eq:single_diagram_normalisation}
\mathcal{N} = N_c^{\left(2 N_{q_I} + N_{q_E}\right)/2} (N_c^2-1)^{\left(2 N_{g_I} + N_{g_E}\right)/2}.
\end{equation}
Note that $\ket{\xi_n}$ does not necessarily have unit norm. In fact, we can see from Eq.~\eqref{eq:vector_xi} that $\braket{\xi_n|\xi_n}$ is the fully-contracted squared colour factor $C_n^2$ for our chosen Feynman diagram $F_n$:
\begin{equation}\label{eq:xi_n_squared}
\braket{\xi_n | \xi_n} = \sum_{
\substack{
a_1, a_2, \ldots \in \{1, \ldots, N_c^2-1\}\\
i_1, i_2, \ldots \in \{1, \ldots, N_c\}\\
j_1, j_2, \ldots \in \{1, \ldots, N_c\}
}
} \left( C^{a_1 a_2 \ldots}_{n, i_1 i_2 \ldots, j_1 j_2 \ldots} \right)^2.
\end{equation}

Equation~\eqref{eq:amplitude_output_state} implies that the probability that the measurement in step~\ref{step:measurement} yields $\ket{\Omega}_{I+\mathcal{U}}$ is:
\begin{equation}\label{eq:probability_omega_measured}
P \left( \ket{\Omega}_{I+\mathcal{U}} \right) =
\frac{1}{\mathcal{N}^2}\braket{\xi_n | \xi_n}.
\end{equation}
Thus, if $C_n^2$ is known beforehand, we can use Eq.~\eqref{eq:probability_omega_measured} to verify that the output state $\ket{\Psi}$ is correctly being produced by performing multiple shots of the circuit, whereby we run the quantum circuit multiple times and count the fraction of times that the measurement yields $\ket{\Omega}_{I+\mathcal{U}}$.
Conversely, Eq.~\eqref{eq:probability_omega_measured} can also be used to estimate $C_n^2$ from the outcome of multiple shots, but while this provides a straight-forward and transparent way to validate the simulation, we emphasise that more sophisticated measurement schemes are available such as quantum amplitude estimation~\cite{Brassard:2000,Grinko:2019,Suzuki:2019,Nakaji:2020}, which compared to a repeated-measurement strategy could provide a quadratic speedup for obtaining $C_n^2$ at a given level of precision.

\section{Interference of multiple Feynman diagrams}\label{sec:interference}

\begin{figure}[]
\centering
\includegraphics[scale=1.04]{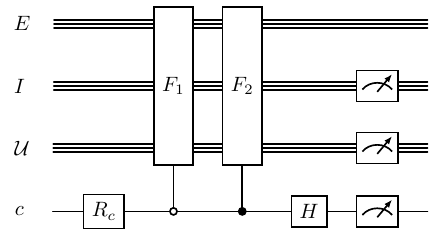}
\caption{Quantum circuit to compute the sum and interference of two Feynman diagrams $F_1$ and $F_2$.
Qubits are grouped into registers, shown as triple horizontal lines.
Registers $E$ and $I$ encode the states of external and internal particles respectively, while $\mathcal{U}$ is the unitarisation register and $c$ is a single-qubit register (shown as a single horizontal line) used for combining the two Feynman diagrams.
A meter icon represents measurement.
All other boxes depict unitary operations defined in the main text.
White circles or black circles indicate, respectively, controlling on a $\ket{0}$ or $\ket{1}$ state.
}\label{fig:interference_F1_F2}
\end{figure}

Now that we have shown how to simulate the colour part of a single Feynman diagram, in this section we present a method to use the quantum computer to calculate interferences of multiple Feynman diagrams.
Let us start by considering the case where we wish to calculate the sum (and hence the interference) of two diagrams, $F_1$ and $F_2$.
The external particles in the two diagrams are of course the same.
To ensure that the two diagrams require the same number of qubits and share the same normalisation factor Eq.~\eqref{eq:single_diagram_normalisation}, we will for this example also assume that the number of closed quark-loops and internal gluons does not differ between the diagrams, although later in this section we will lift this restriction.
We introduce a new ancilla qubit $c$ and define the following single-qubit rotation operation $R_c$:
\begin{equation}\label{eq:R_c}
R_c = \frac{1}{\sqrt{|K_1|^2 + |K_2|^2}}
\begin{pmatrix}
K_1 & -K_2 \\
K_2 & K_1
\end{pmatrix}.
\end{equation}
In the present algorithm, the kinematic factors $K_1$ and $K_2$ need to have been classically computed in advance for a chosen kinematic point and spin configuration, but in the future one could incorporate a full quantum simulation of the kinematics.
By running the circuit shown in Fig.~\ref{fig:interference_F1_F2}, the amplitude in Eq.~\eqref{eq:ampl_sum_diagrams} is obtained via a sequence of states that we will now define in detail.

At the start of the circuit, each register $r$ is in the reference state $\ket{\Omega}_r \equiv \ket{0\ldots 0}_r$ and so the state of the quantum computer is
\begin{equation}\label{eq:interference_F1_F2_initialstate}
\ket{\Omega}_E \ket{\Omega}_I \ket{\Omega}_{\mathcal{U}} \ket{\Omega}_c.
\end{equation}
After applying the $R_c$ gate defined in Eq.~\eqref{eq:R_c} to the ancilla qubit $c$, the state of the quantum computer becomes
\begin{equation}\label{eq:interference_F1_F2_state_after_Rc}
\ket{\Omega}_E \ket{\Omega}_I \ket{\Omega}_{\mathcal{U}} 
\left(\frac{K_1 \ket{0}_c + K_2 \ket{1}_c}{\sqrt{|K_1|^2 + |K_2|^2}}\right).
\end{equation}
A controlled-$F_1$ operation is then performed on registers $E, I, \mathcal{U}$, where the control is on qubit $c$ being in the $\ket{0}$ state.
Here for brevity we will abuse notation by using $F_n$ not only to denote a particular diagram, but also to denote a quantum circuit constructed according to the instructions of Sec.~\ref{sec:ampl_level_tracing} (excluding measurement) to simulate that diagram.
Since the operation is controlled, the state specified in Eq.~\eqref{eq:amplitude_output_state} is produced for the $\ket{0}_c$ component of wavefunction while leaving the $\ket{1}_c$ component unchanged, thus creating the following state:
\begin{equation}
\frac{\ket{\Omega}_I\ket{\Omega}_{\mathcal{U}}}{\sqrt{|K_1|^2 + |K_2|^2}}
\left(K_1 \frac{\ket{\xi_1}_E}{\mathcal{N}} \ket{0}_c + K_2 \ket{\Omega}_E \ket{1}_c \right)
+ \left( \textrm{terms orthogonal to } \ket{\Omega}_I\ket{\Omega}_{\mathcal{U}} \right).
\end{equation}
Next, a controlled-$F_2$ operation is performed, this time controlled on qubit $c$ being in the $\ket{1}$ state, thereby producing the following state:
\begin{equation}
\ket{\Omega}_I\ket{\Omega}_{\mathcal{U}} \left(\frac{K_1 \ket{\xi_1}_E \ket{0}_c + K_2 \ket{\xi_2}_E \ket{1}_c}{\mathcal{N} \sqrt{|K_1|^2 + |K_2|^2}}\right)
+ \big( \textrm{terms orthogonal to } \ket{\Omega}_I\ket{\Omega}_{\mathcal{U}} \big).
\end{equation}
Finally, a Hadamard gate
\begin{equation}\label{eq:hadamard_gate}
H = \frac{1}{\sqrt{2}}\begin{pmatrix}
1 & 1 \\
1 & -1
\end{pmatrix}
\end{equation}
is applied to qubit $c$, causing the state of the quantum computer to become
\begin{equation}\label{eq:output_state_F1_F2}
\frac{1}{\mathcal{N}\sqrt{2}}
\left(\frac{K_1 \ket{\xi_1}_E + K_2 \ket{\xi_2}_E}{\sqrt{|K_1|^2 + |K_2|^2}} \right) \ket{\Omega}_I\ket{\Omega}_\mathcal{U}\ket{\Omega}_c
+ \big(\textrm{terms orthogonal to } \ket{\Omega}_I \ket{\Omega}_\mathcal{U} \ket{\Omega}_c \big).
\end{equation}
The $\ket{\Omega}_I\ket{\Omega}_{\mathcal{U}}\ket{\Omega}_c$ component of this output state encodes the amplitude $\mathcal{A}$, as can be seen by comparing to Eq.~\eqref{eq:ampl_sum_diagrams} and recalling that $\ket{\xi_n}$ encodes the colour factor $C_n$ of diagram $F_n$ via Eq.~\eqref{eq:vector_xi}.
It follows from Eq.~\eqref{eq:output_state_F1_F2} that if the $I$, $\mathcal{U}$, and $c$ registers are measured, the probability of obtaining $\ket{\Omega}_I\ket{\Omega}_{\mathcal{U}}\ket{\Omega}_c$ will be
\begin{equation}\label{eq:P_omega_F1_F2}
P\big(\ket{\Omega}_I\ket{\Omega}_{\mathcal{U}}\ket{\Omega}_c\big) = \frac{
|K_1|^2 \braket{\xi_1|\xi_1}+
K_1^{*} K_2 \braket{\xi_1|\xi_2}+
K_2^{*} K_1 \braket{\xi_2|\xi_1}+
|K_2|^2 \braket{\xi_2|\xi_2}
}{2 \left(|K_1|^2 + |K_2|^2\right) \mathcal{N}^2}.
\end{equation}
By examining Eqs.~\eqref{eq:ampl_sum_diagrams} and~\eqref{eq:vector_xi}, it can be seen that the numerator of the RHS of Eq.~\eqref{eq:P_omega_F1_F2} is in fact the squared amplitude $|\mathcal{A}|^2$, including all interference effects between the two diagrams.

The above procedure can be generalised to an arbitrary number $N_D$ of Feynman diagrams $F_0, \ldots, F_{N_D-1}$.
To do so, the single ancilla qubit $c$ is replaced by a larger ancilla register $D$ of sufficient size to have at least $N_D$ basis states.
This means $D$ contains $\lceil \log_2(N_D) \rceil$ qubits, with basis states $\ket{0}, \ket{1}, \ldots, \ket{2^{\lceil \log_2(N_D) \rceil}-1}$.
For simplicity, let us for now continue to assume that each Feynman diagram has the same numbers of internal particles, so that the internal particle register $I$ required for each diagram is the same.
The $R_c$ gate is replaced by a gate $R_D$ defined such that
\begin{equation}\label{eq:R_D}
R_D \ket{\Omega}_D = \frac{1}{\left(\sum_m |K_m|^2\right)^{1/2}} \sum_{n=0}^{N_D-1} K_n \ket{n}_D.
\end{equation}
In the special case where $N_D$ is a power of $2$ and where all $K_n=1$, the $R_D$ operation can be implemented by simply applying an $H$ gate to each qubit of the $D$ register,
while in general an $R_D$ gate achieving Eq.~\eqref{eq:R_D} can straight-forwardly be constructed by classically pre-computing all $K_n$ at a chosen kinematic point and then employing a sequence of single-qubit rotations and \texttt{CNOT} operations.
The two controlled-$F_n$ operations in Fig.~\ref{fig:interference_F1_F2}, are generalised to a sequence of $N_D$ controlled-$F_n$ operations, with the $k$\textsuperscript{th} operation applying $F_k$ to registers $E, I, \mathcal{U}$, conditioned on the $D$ register being in the state $\ket{k}$:
\begin{equation}
\prod_{k=0}^{N_D-1} \left[
\big(\ket{k}\bra{k}\big)_D F_n
+ \left(\id_D-\big(\ket{k}\bra{k}\big)_D\right) \id_E \id_I \id_{\mathcal{U}}
\right],
\end{equation}
where $\id_r$ denotes the identity operator acting on register $r$.
Finally, the $H$ gate in Fig.~\ref{fig:interference_F1_F2} is replaced by a gate $S_D$ acting on a general state $\ket{\psi}_D$ of the $D$ register in the following way:
\begin{equation}\label{eq:generalised_sum_N}
S_D \ket{\psi}_D = \frac{1}{\sqrt{N_D}} \left( \sum_{n=0}^{N_D-1} \braket{n|\psi}\right)  \ket{\Omega}_D
+ \left(\textrm{terms orthogonal to } \ket{\Omega}_D \right).
\end{equation}
In matrix form, the $S_D$ gate is simply a $2^{\lceil \log_2(N_D) \rceil}$-by-$2^{\lceil \log_2(N_D) \rceil}$ matrix in which the first $N_D$ entries of the first row are $1/\sqrt{N_D}$, any remaining entries of the first row are zero, and the remaining rows can be arbitrary.
For the special case where $N_D$ is a power of $2$, Eq.~\eqref{eq:generalised_sum_N} can be achieved by simply applying an $H$ gate to each qubit of the $D$ register, whereas again the general case is implemented by a suitable sequence rotations and \texttt{CNOT} gates.
Note that unlike $R_D$, the $S_D$ gate is independent of kinematics.

The output state from running this circuit is
\begin{equation}\label{eq:interfered_output_state}
\frac{1}{\mathcal{N}_{N_D}}
\left(\sum_{n=0}^{N_D-1} K_n \ket{\xi_n}_E \right) \ket{\Omega}_{I+\mathcal{U}+D} + (\textrm{terms orthogonal to }\ket{\Omega}_{I+\mathcal{U}+D}).
\end{equation}
where we have defined $\mathcal{N}_{N_D} = \mathcal{N} \sqrt{N_D} \sqrt{\sum_m{|K_m|^2}}$.
This state, analogously to Eq.~\eqref{eq:output_state_F1_F2}, encodes the scattering amplitude from Eq.~\eqref{eq:ampl_sum_diagrams}.
It follows from Eq.~\eqref{eq:interfered_output_state} that if the $I$, $\mathcal{U}$, and $D$ registers are measured, the state $\ket{\Omega}_I\ket{\Omega}_{\mathcal{U}}\ket{\Omega}_D$ will be obtained with probability
\begin{equation}\label{eq:P_omega_interference}
P\left( \ket{\Omega}_I\ket{\Omega}_{\mathcal{U}}\ket{\Omega}_D \right) =
\frac{1}{\mathcal{N}_{N_D}^2}
\left| \sum_n K_n \ket{\xi_n}_E \right|^2.
\end{equation}
By examining Eqs.~\eqref{eq:ampl_sum_diagrams} and~\eqref{eq:vector_xi}, it can be seen that up to an overall normalisation factor, the RHS of Eq.~\eqref{eq:P_omega_interference} is the squared amplitude $|\mathcal{A}|^2$ for the chosen kinematic point and spin configuration, with all interference effects between the diagrams included.

So far we have assumed that the number of internal gluons and closed quark lines does not vary between diagrams, so that all diagrams can be implemented using the same internal-particle register $I$.
If this assumption does not hold, the above algorithm to sum and interfere Feynman diagrams can still be employed, but one must be aware that the normalisation factor $\mathcal{N}$ varies with the number of internal particles in the diagram according to Eq.~\eqref{eq:single_diagram_normalisation}, which would introduce a diagram-dependent weighting in Eq.~\eqref{eq:interfered_output_state}.
This unphysical weighting can, however, easily be removed by re-scaling the diagram's kinematic factor $K_n$ by the amount necessary to cancel it.

It is interesting to notice that the RHS of Eq.~\eqref{eq:P_omega_interference} does not necessarily have any particular scaling with the number of Feynman diagrams, since the growth of the denominator factor on the RHS of Eq.~\eqref{eq:P_omega_interference} can get compensated, partly or entirely, by the growth of the numerator.
Indeed, it is easy to see that if all the Feynman diagrams had identical values, the numerator and denominator would both scale as $|K|^2 N_D^2$, where $K$ is the kinematic factor of each diagram, and so Eq.~\eqref{eq:P_omega_interference} would be invariant under changes of the number of diagrams.
In an opposite extreme case where one diagram is non-zero and all remaining diagrams turn out to be zero, the probability in Eq.~\eqref{eq:P_omega_interference} would scale as $1/N_D$, and so the number of shots required to obtain the amplitude at a given level of precision would scale as $N_D$, which is the same scaling as a classical computation (assuming that neither the classical computation nor the quantum computation exploits the knowledge that most diagrams are zero).
Even in this second extreme case, quantum amplitude estimation~\cite{Brassard:2000,Grinko:2019,Suzuki:2019,Nakaji:2020} techniques could produce a quadratic speedup.
Realistic processes would lie in between these two cases, and so in general a quantum simulation would scale favourably with $N_D$ in comparison to a classical computation of the diagrams.

The circuit depth is also expected to scale well with the number of Feynman diagrams, especially if one exploits shared structures between diagrams.
In the present work we explore two ways to do this, firstly by factoring out shared sub-diagrams, which is illustrated in the example shortly to be discussed, and secondly by exploiting exchange symmetries, which is the subject of Sec.~\ref{sec:permutation}.

\begin{figure}[]
\center
        \begin{subfigure}{0.45\textwidth}
                 \includegraphics[width=0.99\textwidth]{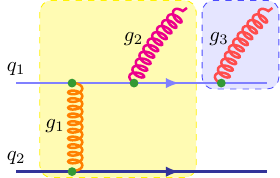}
        \end{subfigure}
\hfill
        \begin{subfigure}{0.45\textwidth}
                 \includegraphics[width=0.99\textwidth]{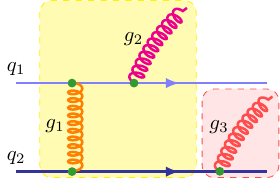}
        \end{subfigure}
        \caption{\label{fig:AB}%
                Feynman diagrams $F_A$ (left) and $F_B$ (right).}
\end{figure}
\begin{figure}[]
\center
\includegraphics[width=0.9\textwidth]{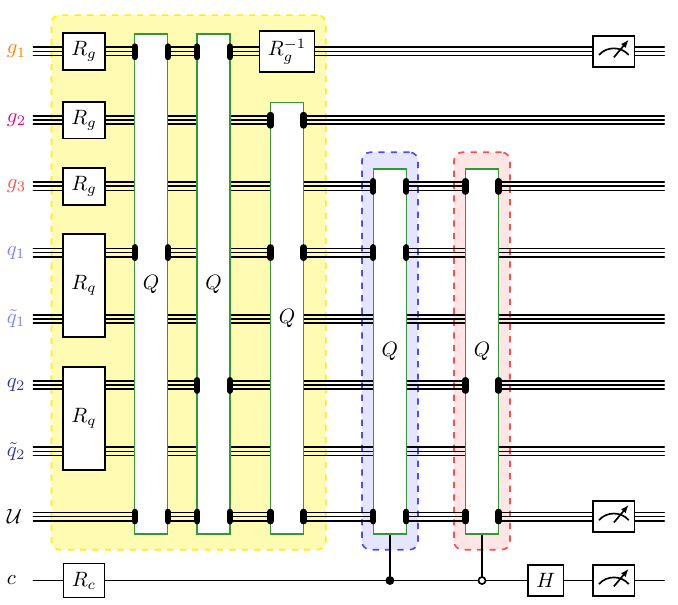}
        \caption{\label{fig:ABcircuit}%
Quantum circuit to calculate the sum and interference of the two Feynman diagrams $F_A$ and $F_B$ shown in Fig.~\ref{fig:AB} and obtain their sum and interference.
The parts shared between the two Feynman diagrams are implemented once, by the section of the circuit shaded in yellow.
The part where the diagrams differ, \textit{i.e.} the emission of $g_3$, is implemented by the blue shaded section for diagram $F_A$ and the red shaded section for diagram $F_B$.
The output of the circuit encodes the fully-interfered squared sum $|F_A + F_B|^2$ of the two diagrams.
}
\end{figure}

To validate our method for calculating interferences, we now apply it to two Feynman diagrams, $F_A$ and $F_B$, which are shown in Fig.~\ref{fig:AB}.
For simplicity, we set both kinematic factors $K_A = K_B = 1$.
The sum and interference of $F_A$ and $F_B$ can be obtained using the quantum circuit shown in Fig.~\ref{fig:ABcircuit}.
We employ a convenient optimisation to reduce the circuit depth: noting that the two diagrams in Fig.~\ref{fig:AB} share a sub-diagram, which is highlighted in yellow, the circuit in Fig.~\ref{fig:ABcircuit} factors the shared sub-diagram out of the controlled-$F_n$ operations that are prescribed for generic pairs of diagrams in Fig.~\ref{fig:interference_F1_F2}.
The shared sub-diagram is therefore only implemented once in the circuit, and is not controlled by the $c$ qubit.
This circuit has been implemented in {\sc Qiskit} and run on a noiseless quantum emulator for $10^7$ shots.
By inverting Eq.~\eqref{eq:P_omega_F1_F2} and inserting $K_1=K_2=1$ as well as $\mathcal{N}^2=N_c^2 \left(N_c^2 -1 \right)^4$, with $N_c=3$, we can infer from the simulation that the squared sum of the diagrams, summed over colours, is:
\begin{align}
\label{eq:exp}
|F_A+F_B|^2 = 4\, \mathcal{N}^2 P\big(\ket{\Omega}_I\ket{\Omega}_{\mathcal{U}}\ket{\Omega}_c\big) = {}& 7.65 \pm 0.34.
%P\big(\ket{\Omega}_I\ket{\Omega}_{\mathcal{U}}\ket{\Omega}_c\big) = 5.19\times 10^{-5} -> 7.65297,
\end{align}
This is consistent with the theoretical value for the squared sum of the Feynman diagrams with $K_A$ and $K_B$ both set to 1:
\begin{align}
|F_A+F_B|^2 = \frac{2 \left(N_c^6-3 N_c^4+5 N_c^2-3\right)
   T_F^4}{N_c^2} = \frac{22}{3} \approx 7.33.
\end{align}

\section{Feynman diagrams with identical particles}\label{sec:permutation}
Having shown in Sec.~\ref{sec:interference} how to obtain the sums and interferences of generic Feynman diagrams, in this section we showcase the value of our methodology by presenting a particularly efficient way to obtain the sums and interferences of Feynman diagrams where some of the external particles are identical.
In considering this class of processes, we find hints of a potential quantum advantage.

\subsection{2-gluon example}\label{sec:2gluonpermuation}
\begin{figure}[]
\centering
\vspace{0.25cm}
\includegraphics[]{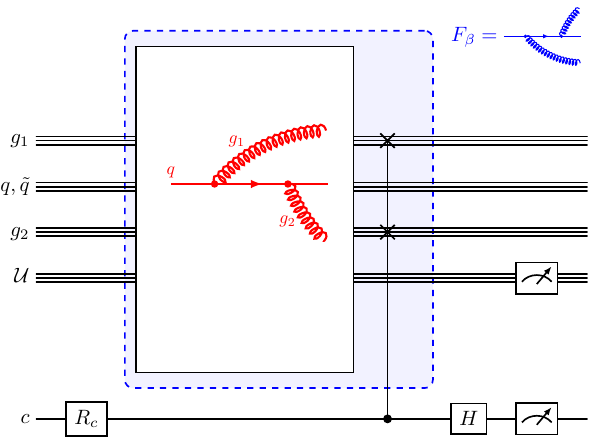}
\caption{Efficient quantum circuit to simulate two Feynman diagrams $F_{\alpha}$ (red) and $F_{\beta}$ (blue) and obtain their sum and interference.
$F_{\alpha}$ is simulated directly whereas $F_{\beta}$ is obtained from $F_{\alpha}$ by swapping the states of the two gluons $g_1$ and $g_2$.
Registers~$q$ and~$\tilde{q}$ encode the outgoing and incoming state of the quark line, respectively.
The solid vertical line with two $\times$ symbols represents a swap of the gluon states, controlled on qubit $c$ being in the state $\ket{1}$.
The output of the circuit encodes $|F_{\alpha} + F_{\beta}|^2$ with full interference effects.
}\label{fig:ggEmissionCircuits}
\end{figure}

As a simple example, let us start by considering in this sub-section the emission of two gluons from a single quark line.
\begin{figure}
     \centering
     \begin{subfigure}[b]{0.49\textwidth}
     \centering
     \includegraphics[width=0.6\textwidth]{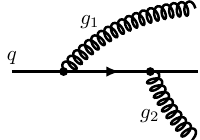}
     \vspace{0.25cm}
     \caption{\centering Feynman diagram $F_{\alpha}$}
     \label{fig:sublabel}
     \end{subfigure}
     \begin{subfigure}[b]{0.49\textwidth}
     \centering
     \includegraphics[width=0.6\textwidth]{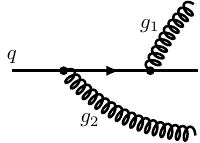}
     \vspace{0.25cm}
     \caption{\centering Feynman diagram $F_{\beta}$}
     \label{fig:sublabel2}
     \end{subfigure}
     \vspace{0.5cm}
        \caption{Feynman diagrams for two sequential gluon emissions from a quark line.}
        \label{fig:feynman_diagrams_doubleemission}
\end{figure}
We will consider two of the Feynman diagrams contributing to this process, which are shown in Fig.~\ref{fig:feynman_diagrams_doubleemission}.
In principle we could directly apply the method from Sec.~\ref{sec:interference} to implement the sum (and hence the interference) of these diagrams.
However, since the second diagram $F_\beta$ in Fig.~\ref{fig:feynman_diagrams_doubleemission} is related to the first diagram $F_\alpha$ by simply relabelling the two gluons, a far more efficient implementation can be achieved, which is shown in Fig.~\ref{fig:ggEmissionCircuits} and will be explained step-by-step below.
The key observation is that $F_\beta$ can be simulated by implementing $F_\alpha$ followed by a swap of the two gluon registers.
This means $F_\alpha$ can be factored out of the two controlled-swap operations required in Fig.~\ref{fig:interference_F1_F2} and performed without being controlled by the $c$ qubit, so that only the swap operation needs to be controlled by the $c$ qubit.
Exploiting the exchange symmetry of the gluons in this way has several benefits: the resulting circuit contains only 2 $Q$ gates instead of 4, and requires far fewer controlled operations, both of which properties reduce the circuit depth.
By including more registers to encode the states of additional particles, the circuit in Fig.~\ref{fig:ggEmissionCircuits} could be used to calculate the sum and interference of any two Feynman diagrams in which the final state contains a pair of gluons.

We will now provide a step-by-step account of the detailed operation of the circuit in Fig.~\ref{fig:ggEmissionCircuits}.
Just as in Sec.~\ref{sec:interference}, the wavefunction is initially as shown in Eq.~\eqref{eq:interference_F1_F2_initialstate}, although in this particular example there is no $I$ register since the diagrams in Fig.~\ref{fig:feynman_diagrams_doubleemission} contain no internal gluons or closed quark loops and so the initial state of the quantum computer is simply:
\begin{equation}
\ket{\Omega}_E \ket{\Omega}_{\mathcal{U}} \ket{\Omega}_c.
\end{equation}
For this example, the $E$ register is composed of two registers $g_1$ and $g_2$ which represent the states of the two gluons, and a pair of registers $q$ and $\tilde{q}$ to represent, respectively, the fundamental and antifundamental $\mathfrak{su}(N_c)$ indices encoding the colour of the quark line, as specified in Sec.~\ref{sec:ampl_level_tracing}.
We can therefore explicitly write the initial state of the $E$ register as follows:
\begin{equation}
\ket{\Omega}_E = \ket{\Omega}_{g_1} \ket{\Omega}_{g_2} \ket{\Omega}_{q} \ket{\Omega}_{\tilde{q}}.
\end{equation}
After applying the $R_c$ gate from Eq.~\eqref{eq:R_c} with $K_1 = K_\alpha$ and $K_2 = K_\beta$, we obtain the state specified in Eq.~\eqref{eq:interference_F1_F2_state_after_Rc}, or more explicitly for the present example:
\begin{equation}
\ket{\Omega}_{g_1} \ket{\Omega}_{g_2} \ket{\Omega}_{q} \ket{\Omega}_{\tilde{q}} \ket{\Omega}_{\mathcal{U}} 
\left(\frac{K_\alpha \ket{0}_c + K_\beta \ket{1}_c}{\sqrt{|K_\alpha|^2 + |K_\beta|^2}}\right).
\end{equation}
After simulating $F_\alpha$, we obtain according to Eq.~\eqref{eq:amplitude_output_state} the state
\begin{equation}\label{eq:state_after_FA}
\frac{\ket{\xi_\alpha}_E}{\mathcal{N}}\ket{\Omega}_{\mathcal{U}}
\left(\frac{K_\alpha \ket{0}_c + K_\beta \ket{1}_c}{\sqrt{|K_\alpha|^2 + |K_\beta|^2}}\right)
+ \big(\textrm{terms orthogonal to }\ket{\Omega}_{\mathcal{U}} \big),
\end{equation}
where
\begin{equation}\label{eq:xi_A}
\ket{\xi_\alpha}_E = \sum_{a,b,i,j,l} T^b_{il} T^a_{lj} \ket{a}_{g_1} \ket{b}_{g_2} \ket{i}_{q} \ket{j}_{\tilde{q}}
\end{equation}
encodes the colour factor $C_\alpha = \sum_l T^b_{il} T^a_{lj}$ of $F_\alpha$.
We next apply a controlled-swap gate \texttt{CSWAP}, which in general acts on a control qubit $q_0$ and two equal-sized particle registers $r_1$ and $r_2$ in the following way:
\begin{equation}
\texttt{CSWAP} \left[ \left( \alpha\ket{0}_{q_0} + \beta\ket{1}_{q_0} \right) \ket{\psi_1}_{r_1} \ket{\psi_2}_{r_2} \right] = \alpha \ket{0}_{q_0} \ket{\psi_1}_{r_1} \ket{\psi_2}_{r_2} + \beta \ket{1}_{q_0} \ket{\psi_2}_{r_1} \ket{\psi_1}_{r_2}.
\end{equation}
Noting that swapping the state of the $g_1$ and $g_2$ registers in $\ket{\xi_\alpha}$ produces the state
\begin{align}
\ket{\xi_\beta}_E &= \sum_{a,b,i,j,l} T^b_{il} T^a_{lj} \ket{b}_{g_1} \ket{a}_{g_2} \ket{i}_{q} \ket{j}_{\tilde{q}} \\
 &= \sum_{a,b,i,j,l} T^a_{il} T^b_{lj} \ket{a}_{g_1} \ket{b}_{g_2} \ket{i}_{q} \ket{j}_{\tilde{q}},
\end{align}
which encodes the colour factor $C_\beta = \sum_l T^a_{il} T^b_{lj}$ of the Feynman diagram $F_\beta$, it can be seen that applying the controlled-swap operation specified in Fig.~\ref{fig:ggEmissionCircuits} to the state in Eq.~\eqref{eq:state_after_FA} produces the state
\begin{equation}
\frac{1}{\mathcal{N}} \sum_{a,b,i,j,l}
\left(\frac{K_\alpha T^b_{il} T^a_{lj} \ket{0}_c + K_\beta T^a_{il} T^b_{lj} \ket{1}_c}{\sqrt{|K_\alpha|^2 + |K_\beta|^2}}\right)
\ket{a}_{g_1} \ket{b}_{g_2} \ket{i}_{q} \ket{j}_{\tilde{q}} \ket{\Omega}_{\mathcal{U}} 
+ \big(\textrm{terms orthogonal to } \ket{\Omega}_{\mathcal{U}} \big).
\end{equation}
Lastly, applying the $H$ gate defined in Eq.~\eqref{eq:hadamard_gate} to the $c$ qubit, the final output state of the circuit in Fig.~\ref{fig:ggEmissionCircuits} is found to be
\begin{equation}\label{eq:output_state_gg_permutation}
\frac{1}{\mathcal{N}\sqrt{2}} \sum_{a,b,i,j,l}
\left(\frac{K_\alpha T^b_{il} T^a_{lj} + K_\beta T^a_{il} T^b_{lj}}{\sqrt{|K_\alpha|^2 + |K_\beta|^2}}\right)
\ket{a}_{g_1} \ket{b}_{g_2} \ket{i}_{q} \ket{j}_{\tilde{q}} \ket{\Omega}_{\mathcal{U}} \ket{\Omega}_c
+ \big(\textrm{terms orthogonal to } \ket{\Omega}_{\mathcal{U}}\ket{\Omega}_c \big),
\end{equation}
which encodes the sum $K_\alpha C_\alpha + K_\beta C_\beta$ of the two Feynman diagrams as desired.
It follows from Eq.~\eqref{eq:output_state_gg_permutation} that if the ${\mathcal{U}}$ and $c$ registers are then measured, the probability of obtaining $\ket{\Omega}_{\mathcal{U}} \ket{\Omega}_c$ is
\begin{equation}
\label{eq:probabilityFAFBswap}
P \left( \ket{\Omega}_{\mathcal{U}} \ket{\Omega}_c \right) =
\frac{1}{2\mathcal{N}^2} \frac{\sum_{a,b} (|K_\alpha|^2 + |K_\beta|^2) \textrm{Tr}(T^a T^b T^b T^a) + 2\textrm{Re}(K_\alpha^* K_\beta) \textrm{Tr}(T^a T^b T^a T^b)}{|K_\alpha|^2 + |K_\beta|^2}
\end{equation}
where the numerator is the squared sum $|F_\alpha + F_\beta|^2$ of the two Feynman diagrams, including all interference effects.

As a validation, we implemented this example using \textsc{Qiskit} and ran $10^7$ shots of the circuit on a noiseless quantum emulator, setting $K_\alpha=K_\beta=1$ for simplicity.
By inverting Eq.~\eqref{eq:probabilityFAFBswap} and setting $\mathcal{N}^2=N_c \left(N_c^2-1 \right)^2$ where $N_c=3$, our simulation gives the following estimate for the squared sum of the Feynman diagrams at the chosen kinematic point:
\begin{align}
 |F_\alpha + F_\beta|^2 = 4\, \mathcal{N}^2 P\left( \ket{\Omega}_{\mathcal{U}} \ket{\Omega}_c \right) =  9.357 \pm 0.027.
% 9.35747
% P\left( \ket{\Omega}_{\mathcal{U}} \ket{\Omega}_c \right) = 0.0121842
\end{align}
Noting that the theoretical values for the colour-summed colour parts of the squared Feynman diagrams and their interference read
\begin{align}
 |C_\alpha|^2 = |C_\beta|^2 = \frac{(N_c^2-1)^2}{4N_c} \quad \textrm{and} \quad  C_\alpha C_\beta^* = \frac{1-N_c^2}{4N_c} ,
\end{align}
we see that the results of the simulation at the chosen kinematic point are well in agreement with the analytic expectation of
\begin{equation}
|F_\alpha + F_\beta|^2 =  \frac{N_c^4-3N_c^2+2}{2N_c} = \frac{28}{3} \approx 9.333.
% P\left( \ket{\Omega}_U \ket{\Omega}_c \right) = 0.0121528
\end{equation}
In general, such a swap-based technique can be employed to more efficiently implement the interference of Feynman diagrams whenever the external states of a process include a pair of identical particles.

\subsection{Generalisation to many identical particles}\label{sec:permutations_N}

The technique from Section~\ref{sec:2gluonpermuation} can be generalised to the case of $N$ identical particles.
Given a Feynman diagram $F$ containing $N$ identical external particles, there exists a set of $N!$ diagrams related to $F$ by permutations of the identical particles.
We focus on the general case where $F$ itself has no graph symmetries under such permutations, and so the $N!$ permuted diagrams are mutually distinct.
Remarkably, the sum (and hence interference) of these $N!$ diagrams can be obtained by implementing a single diagram followed by a small number $N_\textrm{swaps}$ of \texttt{CSWAP} operations, where $N_\textrm{swaps}$ will be quantified below after we have described our quantum circuit construction.
The single $c$ qubit in Fig.~\ref{fig:ggEmissionCircuits} is replaced by a larger register $\mathcal{P}$ comprising $N_\textrm{swaps}$ qubits, each acting as the control for a distinct \texttt{CSWAP}.
In a procedure building in part on Ref.~\cite{Abrams:1997gk}, we apply \texttt{CSWAP} operations to the $N$ registers encoding the states of the particles, these operations being applied in a layout that corresponds to an $N$-wire sorting network~\cite{US3029413}.
Since the swap operations form a sorting network, any particular permutation $\sigma$ of the $N$ particles can be generated by an appropriate subset of the swap operations, although often there will exist several distinct subsets that generate the same permutation.
To ensure that each permutation is obtained exactly once, the $R_c$ and $H$ gates in Fig.~\ref{fig:ggEmissionCircuits} must be replaced by generalised gates that rotate the control qubits between the reference state $\ket{\Omega}$ and a superposition of $N!$ basis states judiciously chosen to ensure that the \texttt{CSWAP} operations in a particular sorting network generate all $N!$ permutations with no double counting.
As in the previous sections, the kinematic factors $K_n$ can be pre-computed and encoded in the first of the two generalised gates.
Having implemented the circuit for $F$ together with a sorting network and generalised rotations, the final state of the quantum computer will, similarly to Eq.~\eqref{eq:interfered_output_state}, be
\begin{equation}\label{eq:permuted_output_state}
\frac{1}{\mathcal{N}_{N_D}}
\left(\sum_\sigma K_\sigma \ket{\xi_\sigma}_E \right) \ket{\Omega}_{I+{\mathcal{U}}+\mathcal{P}} + (\textrm{terms orthogonal to }\ket{\Omega}_{I+{\mathcal{U}}+\mathcal{P}}),
\end{equation}
where $N_D = N!$, and where $K_\sigma$ and $\ket{\xi_\sigma}$ encode respectively the kinematic and colour factor of the Feynman diagram $F_\sigma$ obtained by applying the permutation $\sigma$ to the $N$ identical particles in the original Feynman diagram $F$.

Sorting networks are a well-studied topic in computer science and so there are many choices of networks available.
Formally, an AKS network~\cite{10.1145/800061.808726} could allow the permutations to be performed with $N_\textrm{swaps} \sim \mathcal{O}\left(N \log N \right)$ \texttt{CSWAP} operations arranged in a circuit of depth $\mathcal{O}(\log(N))$, which is doubly-exponentially shorter than implementing each of the $N!$ diagrams sequentially.
AKS networks have a large constant overhead, but more practical algorithms exist requiring a depth of $\mathcal{O}(N\log(N))$ operations, which is still exponentially shorter than implementing each diagram sequentially.
For small values of $N$, optimal circuits are explicitly known.
For example, given a circuit to implement a single Feynman diagram containing $N=10$ identical particles, one can use 31 $\texttt{CSWAP}$ operations, parallelised over a depth of 7, to generate the sum and interference of all $10! = 3,628,800$ permuted diagrams.
Depth-optimal networks are known for $N \leq 17$ (e.g. depth 10 for $N=17$ as shown in Ref.~\cite{CODISH2019184}) and size-optimal networks are known for $N \leq 12$ (e.g. $N_\textrm{swaps} = 39$ for $N=12$ as shown in Ref.~\cite{harder2022answerbosenelsonsortingproblem}).

\begin{figure}[]
\center
                 \includegraphics[width=0.6\textwidth]{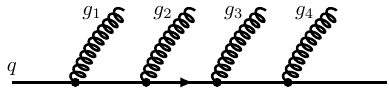}
        \caption{\label{fig:4gluons_FeynmanDiagram}%
                Example Feynman diagram for the sequential emission of four gluons from a quark line.}
\end{figure}
\begin{figure}[]
\vspace{1cm}
\center
                 \includegraphics[width=1.0\textwidth]{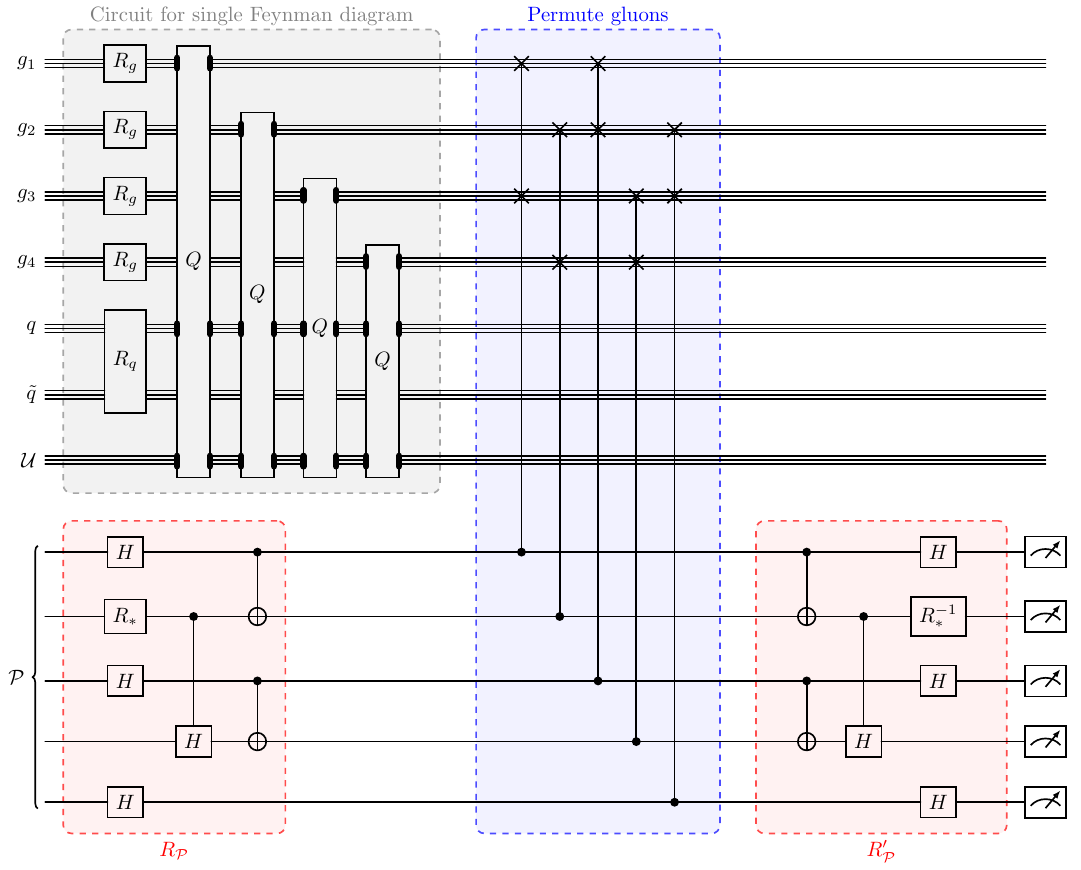}
        \caption{\label{fig:4gluons_circuit}%
Quantum circuit to obtain the sum and interference of 24 Feynman diagrams describing the sequential emission of 4 gluons from a quark line.
The part highlighted in grey implements a single representative Feynman diagram shown in Fig.~\ref{fig:4gluons_FeynmanDiagram}.
The parts highlighted in red and blue generate all 24 permutations of the original diagram.
The fully-interfered squared sum $|\mathcal{A}|^2 = |\sum_\sigma F_\sigma|^2$ of the diagrams is obtained from the output of the circuit.
}
\end{figure}

As a concrete example of our procedure for generating and interfering permuted diagrams, in Fig.~\ref{fig:4gluons_circuit} we explicitly show a quantum circuit for calculating the squared sum, with full interference effects, of all $(4!) = 24$ Feynman diagrams corresponding to the sequential emission of 4 gluons from a quark line.
One representative diagram, shown in Fig.~\ref{fig:4gluons_FeynmanDiagram}, is directly implemented in the circuit, whereas the remaining 23 diagrams are generated by the 5-swap sorting network that follows it.
The sorting network has depth 3 (since the first two swaps can be performed simultaneously, as can the next two) and size $N_\textrm{swaps}=5$, both of which are optimal when the number of identical particles is 4.
The generalised rotation gates have been labelled $R_\mathcal{P}$ and $R_\mathcal{P}'$.
Their definition includes a single-qubit rotation gate
\begin{equation}
R_* = \begin{pmatrix}
\sqrt{1/3} & -\sqrt{2/3} \\
\sqrt{2/3} & \sqrt{1/3}
\end{pmatrix}.
\end{equation}
For simplicity, all 24 kinematic factors have been set to be 1.
If desired, $R_\mathcal{P}$ could instead be replaced by a rotation gate that incorporates kinematic factors pre-computed at a chosen kinematic point.
In Sec.~\ref{sec:physical_devices} we will implement this circuit on a present-day quantum computer in order to validate our techniques.

It is helpful to distinguish the way the permutation symmetries of diagrams are exploited in this quantum algorithm in comparison to the way they are exploited in classical computing algorithms.
Classical calculations can be performed either analytically or numerically.
In analytic calculations, it is straight-forward to exploit permutation symmetries by calculating a symbolic expression for one Feyman diagram and then permuting its variables to generate other related diagrams.
However, analytically calculating Feynman diagrams requires symbolic manipulations that can be computationally expensive and so direct numerical calculations are often preferable.
In such numerical calculations, however, each of the permuted diagrams must be independently evaluated at the desired phase-space points, and so the permutation symmetries between the diagrams are of limited benefit.
Our direct quantum-computer-based simulation of the Feynman diagrams thus provides additional capabilities to exploit symmetries, going beyond what can be achieved in a classical numerical simulation.
For processes with several identical external particles, such as in the production of multiple hard jets, the factorial reduction in the number of Feynman diagrams needing to be implemented could provide a sizeable benefit.

\section{Validation on physical quantum devices}\label{sec:physical_devices}
\begin{figure}[]
\centering
\includegraphics[scale=1.05]{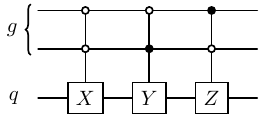}
\caption{Circuit diagram for the $\Qsutwo$ gate acting on a 2-qubit gluon register $g$ and 1-qubit quark register $q$. Boxes labelled $X$, $Y$, and $Z$ represent the three elementary Pauli gates. White and black circles indicate, respectively, controlling on a $\ket{0}$ or $\ket{1}$ state.
}\label{fig:Qsutwo_gate}
\end{figure}

In Sections~\ref{sec:interference} and~\ref{sec:permutation} we included validations of the techniques presented in those sections, by implementing various examples on classically-emulated quantum computers.
In addition, we have also performed validations on a physical quantum computer, as we will present in this section.
Specifically, we ran our two most complicated examples on the 56-qubit trapped-ion quantum computer Quantinuum H2-1.
Details about the properties and performance of the device can be found in Ref.~\cite{PhysRevX.13.041052}.
Quantum computers available today are constrained primarily by noise and decoherence, thus limiting the circuit depths that can be achieved.
In order to operate within these constraints, we perform our physical-device demonstrations using a toy model of QCD where $N_c=2$.
The Lie algebra of the gauge group is thus $\sutwo$, whose fundamental-representation generators $T^a$ (with $a \in \{1,2,3\}$) can be chosen to be unitary%
\footnote{This choice would ordinarily imply $T_F=2$ instead of the conventional $T_F=\frac{1}{2}$ used in the previous sections, but we have corrected for the differences in convention in order to consistently use $T_F=\frac{1}{2}$ throughout this article.}
by defining $T^a = \sigma^a$.
Here $\sigma^a$ are the Pauli matrices, which are readily available as single-qubit gates on many quantum devices.
In $\sutwo$ we can represent fundamental, antifundamental, and adjoint indices by using registers containing, respectively, 1, 1, and 2 qubits.
We can then construct an $\sutwo$ version of the $Q$ gate, denoted $\Qsutwo$, to act on a gluon register $g$ in a colour basis state $\ket{a}$ and a quark register $q$ in a colour basis state $\ket{k}$ in the following way:
\begin{equation}\label{eq:Q_gate_su2}
\Qsutwo\ket{a}_g\ket{j}_q= \sum_{i=1}^2 T^a_{ij} \ket{a}_g\ket{i}_q ,
\end{equation}
where $T^a_{ij}$ now denotes the components of the fundamental-representation generators of $\mathfrak{su}(2)$ instead of $\mathfrak{su}(3)$. Equation~\eqref{eq:Q_gate_su2} is simpler than Eq.~\eqref{eq:Q_gate_behaviour} because here no unitarisation register is required.
An explicit construction of $\Qsutwo$ is shown in Fig.~\ref{fig:Qsutwo_gate}.
We also constructed $\sutwo$ versions of the $R_g$ and $R_q$ gates.

Using this setup, we first implemented the two-diagram interference example from Figs.~\ref{fig:AB} and~\ref{fig:ABcircuit}.
We set the values of the kinematic factors to be $K_A = K_B = 1$.
The circuit was compiled using \textsc{Pytket}~\cite{Sivarajah:2020lfo} version 1.4.0, extension \texttt{pytket-quantinuum 0.44.0}, with setting \texttt{optimization\_level=0}.
By running the circuit for 1,000 shots on the device H2-1, we obtained the result
\begin{equation}
|F_A + F_B|^2 = 1.099 \pm 0.075.
\end{equation}
This is consistent with the analytic expectation
\begin{align}
|F_A+F_B|^2 = \frac{2 \left(N_c^6-3 N_c^4+5 N_c^2-3\right)
   T_F^4}{N_c^2} = \frac{33}{32} \approx 1.031,
\end{align}
where we have set $N_c=2$ and $T_F=\frac{1}{2}$.

To further validate and demonstrate our techniques, we also implemented the 24-diagram interference from Figs.~\ref{fig:4gluons_FeynmanDiagram} and~\ref{fig:4gluons_circuit}.
As in the previous example, we work in $\sutwo$ and set all kinematic factors $K_n=1$.
We compiled the circuit using the \textsc{Pytket} version 2.4.1, extension \texttt{pytket-quantinuum 0.48.0}, with the setting \texttt{optimization\_level=0}.
By running the circuit for 1,000 shots on the device H2-1, we obtained the result
\begin{equation}
\left| \sum_{m=1}^{24} F_{\sigma_{m}} \right|^2 = 11.12 \pm 1.42,
\end{equation}
where $F_{\sigma_1}, F_{\sigma_2}, \ldots, F_{\sigma_{24}}$ are the 24 Feynman diagrams obtained by permuting the ordering of the gluon emissions in Fig.~\ref{fig:4gluons_FeynmanDiagram}.
This result is consistent with the analytic expectation of
\begin{equation}
\left| \sum_{m=1}^{24} F_{\sigma_m} \right|^2 = \frac{24(N_c^8-N_c^6-24N_c^2+24)T_F^4}{N_c^4} = \frac{45}{4} = 11.25,
\end{equation}
where we have set $N_c=2$ and $T_F=\frac{1}{2}$.

\section{Conclusion}
High-energy physics presents a promising domain of application for quantum computers.
In this article we have presented quantum computing techniques to simulate scattering amplitudes in perturbative QCD, which describes the high-energy regime of QCD probed by colliders such as the LHC.
We first showed how the colour parts of individual Feynman diagrams can be encoded into the wavefunction of a quantum computer.
We then showed how the sum of multiple Feynman diagrams can be implemented, providing a natural way to obtain their interferences.
Our methods are well suited to exploiting relations between Feynman diagrams, which we explored in two ways: firstly with diagrams sharing a common sub-diagram, and secondly with diagrams related to one another by permutations of the external particles.
The latter case shows the first hints of a potential quantum advantage, \textit{i.e.}\ indications that quantum computers could outperform classical computers for such calculations.
We validated our techniques by building simulation circuits to obtain the fully-interfered squared sums of Feynman diagrams for several example cases.
The circuits were run on quantum emulators as well as a state-of-the-art 56-qubit trapped-ion quantum computer, and the results were consistent with analytic expectations.

This work opens up several possibilities to run more ambitious applications on present and future physical quantum computers.
This would benefit from tailoring the fundamental implementation of the $Q$ and $G$ gates to suit the architectures of specific present-day devices and could also incorporate a direct quantum simulation of the kinematic parts of the amplitudes.
At the same time, this work also constitutes a conceptual milestone for further developments, with potential future applications ranging from improved parton-shower algorithms to a quantum-accelerated Monte Carlo simulation of scattering cross-sections.

\section*{Acknowledgements}

We thank Zohreh Davoudi, Aida El-Khadra, and Fernando Febres Cordero for helpful discussions.
The work of HC was supported by the U.S. Department of Energy under grant DE-SC0010102.
MP acknowledges support by the German Research Foundation (DFG) through the Research Training Group RTG2044.
This research was supported by the Munich Institute for Astro-, Particle and BioPhysics (MIAPbP), which is funded by the DFG under Germany's Excellence Strategy - EXC-2094 - 390783311.
This work was performed in part at the Aspen Center for Physics, which is supported by National Science Foundation grant PHY-2210452.
This research was supported in part by grant NSF PHY-2309135 to the Kavli Institute for Theoretical Physics (KITP).
This research used resources of the Oak Ridge Leadership Computing Facility, which is a DOE Office of Science User Facility supported under Contract DE-AC05-00OR22725.
The Feynman diagrams were produced using the \textsc{Tikz} package~\cite{tikzManual3-1-10}.
The circuit diagrams were produced using the \textsc{Quantikz} package~\cite{Kay:2018huf}.

\bibliographystyle{utphys.bst}

\bibliography{Qcolour}
\end{document}

%% file: macros.tex
\def\be{\begin{equation}}
\def\ee{\end{equation}}

\newcommand{\alphas}{\ensuremath{\alpha_\text{s}}\xspace}

\newcommand{\newc}{\newcommand}
\newc{\bi}{\begin{itemize}}
\newc{\ei}{\end{itemize}}
\newc{\benu}{\begin{enumerate}}
\newc{\eenu}{\end{enumerate}}
\newc{\bc}{\begin{center}}
\newc{\ec}{\end{center}}
\newc{\bfig}{\begin{figure}}
\newc{\efig}{\end{figure}}
\newc{\qbar}{\bar{q}}
\newc{\go}{\tilde{g}}
\newc{\PB}{\textsc{Powheg-Box}}

\newcolumntype{.}{D{.}{.}{-1}}
\newcolumntype{d}[1]{D{.}{.}{#1}}

\colorlet{tableoverheadcolor}{gray!37.5}
\colorlet{tableheadcolor}{gray!25}
\colorlet{tablerowcolor}{gray!12.5}

\newlength{\width}
\newlength{\height}

\def\draftdate{\relax}
\def\mda{\relax}
\def\mua{\relax}
\def\mla{\relax}
\def\draft{
\def\thtystars{******************************}
\def\sixtystars{\thtystars\thtystars}
\typeout{}
\typeout{\sixtystars**}
\typeout{* Draft mode!
         For final version remove \protect\draft\space in source file *}
\typeout{\sixtystars**}
\typeout{}
\def\draftdate{\today}
\def\mua{\marginpar[\boldmath\hfil$\uparrow$]%
                   {\boldmath$\uparrow$\hfil}\color{black}%
                    \typeout{marginpar: $\uparrow$}\ignorespaces}
\def\mda{\color{red}\marginpar[\boldmath\hfil$\downarrow$]%
                   {\boldmath$\downarrow$\hfil}%
                    \typeout{marginpar: $\downarrow$}\ignorespaces}
\def\mla{\marginpar[\boldmath\hfil$\rightarrow$]%
                   {\boldmath$\leftarrow $\hfil}%
                    \typeout{marginpar: $\leftrightarrow$}\ignorespaces}
\def\Mua{\marginpar[\boldmath\hfil$\Uparrow$]%
                   {\boldmath$\Uparrow$\hfil}\color{black}%
                    \typeout{marginpar: $\uparrow$}\ignorespaces}
\def\Mda{\color{red}\marginpar[\boldmath\hfil$\Downarrow$]%
                   {\boldmath$\Downarrow$\hfil}%
                    \typeout{marginpar: $\downarrow$}\ignorespaces}
\def\Mla{\marginpar[\boldmath\hfil\textcolor{red}{$\Rightarrow$}]%
                   {\boldmath\textcolor{red}{$\Leftarrow $}\hfil}%
                    \typeout{marginpar: $\leftrightarrow$}\ignorespaces}
\overfullrule 5pt
\oddsidemargin 15mm
\marginparwidth 29mm
}